\documentclass[12pt]{article}

\usepackage{graphicx}
\usepackage{epsfig}
\usepackage{amsfonts}
\usepackage{amssymb}
\def\ep{{\epsilon}}

\begin{document}

\title{
Charged isotropic non-Abelian dyonic black branes  }
 \vspace{1.5truecm}
\author{{\large Yves Brihaye,}$^{1}$
{\large Ruben Manvelyan,}$^{2}$
{\large Eugen Radu}$^{3  }$
and {\large D. H. Tchrakian}$^{4,~5}$ 
\vspace*{0.2cm}
\\
$^{1}${\small Physique-Math\'ematique, Universite de
Mons-Hainaut, Mons, Belgium}\\
$^{2}${\small Yerevan Physics Institute, Alikhanian Br. St. 2, 0036 Yerevan, Armenia}\\
$^{3}${\small Departamento de F\'\i sica da Universidade de Aveiro and CIDMA} \\ 
{\small   Campus de Santiago, 3810-183 Aveiro, Portugal}
  \\
$^{4}${\small School of Theoretical Physics -- DIAS, 10 Burlington
Road, Dublin 4, Ireland} 
\\
$^{5}${\small  Department of Computer Science,
National University of Ireland Maynooth,
Maynooth,
Ireland}
} 
\date{\today}
%\pacs{04.20.Jb, 04.40.Nr}
\maketitle
\begin{abstract}

We construct black holes with a Ricci flat horizon in Einstein--Yang-Mills theory 
with a negative cosmological constant,
which approach  asymptotically
an AdS$_d$ spacetime background (with $d\geq 4$).
These solutions are isotropic, $i.e.$ all space directions in a  
hypersurface of constant radial and time coordinates are equivalent,
and possess both electric and magnetic fields. 
We find that the basic properties of the non-Abelian solutions 
are similar to those of the dyonic isotropic branes in 
Einstein-Maxwell theory (which, however, exist in even spacetime dimensions only). 
These black branes possess a nonzero magnetic field strength
on the flat boundary metric,
which leads to a divergent  mass of these solutions, as defined in the usual way.
However, a different picture is found for odd spacetime dimensions,
 where a non-Abelian Chern-Simons term can be incorporated in the action. 
This allows for black brane solutions with a magnetic field
which
vanishes asymptotically. 

\end{abstract}

%%%%%%%%%%%%%%%%%%%%%%%%%%%%%%%%%%%%%%%%%%%%%%%%%%%%%%%%%%%%%%%%%%%%%%%%%%%%%%
%%%%%%                     Introduction
%%%%%%%%%%%%%%%%%%%%%%%%%%%%%%%%%%%%%%%%%%%%%%%%%%%%%%%%%%%%%%%%%%%%%%%%%%%%%%
\noindent{\textbf{~~~Introduction and motivation.--~}}
In recent years there has been some interest in
studying the AdS/CFT correspondence 
\cite{Maldacena:1997re},
\cite{Witten:1998qj},
in the presence of a background magnetic field.  
On the bulk side, this corresponds to solving the 
Einstein-gauge field system of equations,
with suitable boundary conditions such that the AdS background is
approached asymptotically, while
the magnetic field does not trivialize. 
Several new classes of such  solutions have been found in this way,
most of them for the case of main interest of asymptotically 
AdS$_5$ configurations with Abelian fields.
For example, the results in 
\cite{D'Hoker:2009mm},
\cite{D'Hoker:2009bc}
%have 
revealed the existence of a variety of unexpected features
of these solutions;
here we mention
only that their study is relevant for
the issue of the third law of thermodynamics
in the AdS/CFT context.

The investigation of the non-Abelian (nA) generalizations of these solutions  
is only in its beginning stages. 
Considering such configurations is a legitimate task, 
since the gauged supersymmetric models 
generically contain Yang-Mills fields (although usually only Abelian truncations are considered).
To date, the only case investigated systematically 
corresponds to that in four ($d=4$) spacetime dimensions
(see  \cite{Winstanley:2008ac} for a review of these solutions).
The four dimensional  nA asymptotically-AdS (AAdS) solutions
exhibit many new features which are absent for $\Lambda \geq  0$.
For example, stable\footnote{The stability is against linear perturbations, and is not topological.} 
solitons and black holes, possesing a global magnetic charge,
are known to exist in a globally AdS$_4$ background 
even in the absence of a Higgs field 
\cite{Winstanley:1998sn}, \cite{Bjoraker:2000qd}.
However, the results in 
\cite{VanderBij:2001ia},
\cite{Mann:2006jc}  
 show that these Einstein--Yang-Mills (EYM)  black holes solutions have also generalizations 
 with a nonspherical event horizon topology, in particular
with a Ricci flat horizon and a magnetic field which does not vanish asymptotically.
% (see also \cite{Gubser:2008zu}).
They share many of the features of the spherical
configurations in  \cite{Winstanley:1998sn}, \cite{Bjoraker:2000qd}, 
in particular the existence of solutions stable against
linear fluctuations. 
 The only $d>4$ nA AAdS solutions black holes studied more systematically so far are those
possessing spherical event horizon topology \cite{Okuyama:2002mh}-\cite{Brihaye:2009cc}, though
some solutions with Ricci flat horizon have been studied in  
\cite{Manvelyan:2008sv},
\cite{Ammon:2009xh}.
 
In an unexpected development,
the study of the $d=4,5$ EYM  black brane solutions 
 has led to the discovery of holographic superconductors and holographic superfluids, 
 describing condensed phases of strongly coupled, planar, gauge theories
 \cite{Gubser:2008zu}.
Studying such solutions involves the construction
of AAdS electrically charged black  branes, which, below a critical temperature
become unstable to forming YM hair.
However, the magnetic field of these configurations 
vanishes on the boundary, leading to 
a vanishing background magnetic field for the dual theory. 

 The main purpose of this work is to  
present an investigation of $d\geq 4$ 
AAdS isotropic black branes 
  supporting 
both electric and magnetic nA fields.
In contrast to previous studies in the literature,
the magnetic fields  of these solutions  do not vanish on the boundary,
which leads to a variety of interesting features.  
For example,
we find that the mass of these asymptotically AdS solutions, as defined in the usual way,
 always diverges, 
 while the solutions do not posses a regular extremal limit.
   In odd dimensional spacetimes, when a Chern-Simons term is added to the total action, it is found that
a special class of  solutions exhibit a nontrivial magnetic field in the bulk while vanishing asymptotically. 
 
%%%%%%%%%%%%%%%%%%%%%%%%%%%%%%%%%%%%%%%%%%%%%%%%%%%%%%%%%%%%%%%%%%%%%%%%%%%%%%%
%\section{The Einstein--Yang-Mills system}
%%%%%%%%%%%%%%%%%%%%%%%%%%%%%%%%%%%%%%%%%%%%%%%%%%%%%%%%%%%%%%%%%%%%%%%%%%%%%%%
\noindent{\textbf{~~~The Einstein--Yang-Mills system.--~}}
We consider the EYM theory in a $d$-dimensional
 spacetime, with a cosmological constant
$\Lambda=-(d-2)(d-1)/(2L^2)$. The action is
\begin{eqnarray}
\label{action}
I=\int_{\mathcal{M}} d^dx\sqrt{-g}~\left(\frac{1}{16 \pi G}(R-2 \Lambda)-\frac{1}{4}{*F}\wedge F   \right)
+S_{bndy} .
\end{eqnarray}
The boundary terms $S_{bndy}$ include the Gibbons-Hawking term \cite{Gibbons:1976ue} as well as 
%other contributions necessary for a well posed variational principle
the counterterms required for the on-shell action to be finite \cite{Balasubramanian:1999re}.
The Einstein and Yang-Mills equations derived from the above action are
\begin{eqnarray}
\label{EOM}
R_{\mu \nu}-\frac{1}{2}R g_{\mu \nu}+\Lambda g_{\mu \nu} = 8 \pi G~T_{\mu\nu},~~~~~
D_\mu F^{\mu\nu} =0,
\end{eqnarray}
where $D_\mu$ is the gauge derivative and 
the  Yang-Mills stress-energy tensor 
\begin{eqnarray}
\label{Tmunu}
T_{\mu \nu}= \frac{1}{2}\left(F^{IJ}_{\mu\rho}F^{IJ}_{\nu\sigma}g^{\rho\sigma}-\frac{1}{4}g_{\mu\nu} F_{\rho\sigma}^{IJ}F^{IJ\rho\sigma} \right)~.
\end{eqnarray}
 %and $D_\mu=\nabla_\mu+i [A_\mu,~]$ is the gauge derivative.
 
We are interested in static Ricci-flat solutions which approach asymptotically a (planar) AdS$_d$
background.
Also, to simplify the picture, we shall restrict our study to the   following case:
denoting  the radial and time coordinate by $r$ and $t$ respectively and considering
the hypersurfaces  parametrized by $x^i$ ($i=1,\dots, d-2$ and  $r,t$ fixed), we assume that 
all space directions in these hypersurfaces are equivalent.
Thus the field strength and the metric are taken to be invariant under 
space translations and rotations in the planes $(x^i,x^j)$; they are also time independent. 
 Without any loss of generality, a line element with this property 
can be written in the form
\begin{eqnarray}
\label{metric}
ds^2 =g_{rr}(r)dr^2+g_{\Sigma\Sigma}(r) d\Sigma^2_{d-2} +g_{tt}(r) dt^2,
\end{eqnarray}
whith
$d\Sigma^2_{d-2}=(dx^1)^2+\dots +(dx^{d-2})^2 $
the metric on the $(d-2)$-flat space. 
 
  The above symmetry requirements  imply some restrictions on the 
the choice of the gauge group.
Restricting to $SO(n)$ YM fields,
one finds that 
a YM ansatz leading to an isotropic energy-momentum tensor
for $both$ even and odd values of $d$
 is possible
for  $n\geq d+1$ only\footnote{Note that, for even values of $d$,
one can consider instead a gauge 
 group $SO(d-1)$, which leads to isotropic EYM branes.
 A study of this case has been proposed in  \cite{Manvelyan:2008sv} (Ansatz I there).
 However, the properties of those solutions are rather different 
 %as compared 
to the case of interest 
%in this work
here.}. 

In this work we shall consider an  $SO(d+1)$ gauge group, with
 $d(d-1)/2$ $SO(d+1)$ nA gauge fields 
represented by the 1-form
potential $A^{IJ}$ antisymmetric in $I$ and $J$ (with $I,J=1,\dots ,d+1$)
and 
$F^{IJ}=dA^{IJ}+\frac{1}{\hat g}A^{IK}\wedge A^{KJ}$,
with $\hat g$ the Yang-Mills coupling.
Also,  to simplify the relations,
 it is convenient to define
% the coupling constant
\begin{eqnarray}
\label{alpha}
\alpha^2=\frac{4\pi G}{\hat g^2}~.
 \end{eqnarray}

%%%%%%%%%%%%%%%%%%%%%%%%%%%%%%%%%%%%%%%%%%%%%%%%%%%%%%%%%%%%%%%%%%%%%%%%%%%%%%%
%\subsection{Abelian solutions}
%%%%%%%%%%%%%%%%%%%%%%%%%%%%%%%%%%%%%%%%%%%%%%%%%%%%%%%%%%%%%%%%%%%%%%%%%%%%%%%
 \noindent{\textbf{~~~Embedded Abelian solutions.--~}}
 Before proceeding to the non-Abelian case,
 it is instructive to consider the dyonic
 black branes in Einstein-Maxwell theory, ($i.e.$
 the gauge fields taking their values in the $U(1)$ subgroup of $SO(d+1)$).
A gauge field ansatz compatible with the symmetries of the line-element
 (\ref{metric}) can be constructed for an even number of spacetime
 dimensions only, $d=2n+2$ and reads\footnote{The ansatz  
 (\ref{Maxwell-ansatz}), 
 (\ref{metric})
  can be extended to the case of odd $d$ by adding a number of codimensions $y^{\mu}$, with 
  $A_{\mu}^{IJ}=0$;
 however, this leads to anisotropic configurations.
 }
\begin{eqnarray}
\nonumber 
&&
A_1^{IJ}=\frac{w_0^2}{\hat g} x^2 \delta_{[d}^{~I}\delta_{d+1]}^{~J} ,~
A_2^{IJ}=-\frac{w_0^2}{\hat g}x^1\delta_{[d}^{~I}\delta_{d+1]}^{~J},
\\
\nonumber
&&{~~~~~~~~~~~~~~~~~~}\dots,
\\
\label{Maxwell-ansatz}
&&
A_{2n-1}^{IJ}=\frac{w_0^2}{\hat g}x^{2n} \delta_{[d}^{~I}\delta_{d+1]}^{~J},~
A_{2n}^{IJ}=-\frac{w_0^2}{\hat g} x^{2n-1} \delta_{[d}^{~I}\delta_{d+1]}^{~J}, 
\\
\nonumber
&&
%~~~~{\rm and}~~~
A_r^{IJ}=0,~~A_t^{IJ}=\frac{V(r)}{\hat g}\delta_{[d}^{~I}\delta_{d+1]}^{~J},
\end{eqnarray}
with $w_0$ an arbitrary parameter
which fixes the magnetic field in a two plane,
$F_{21}^{IJ}=\dots=F_{2n_{}2n-1}^{IJ}=\frac{2w_0^2}{\hat g}\delta_{[d}^{~I}\delta_{d+1]}^{~J}$.
Choosing a metric gauge with $g_{\Sigma\Sigma}=r^2$,
one finds\footnote{The purely
magnetic limit of this solution, with $Q=0$, has been discussed in \cite{D'Hoker:2009mm}. } a black brane solution with $1/g_{rr}=-g_{tt}=N(r)$,
where
\begin{eqnarray}
\label{metric1}
 N(r)=\frac{r^2}{L^2}-\frac{M_0}{r^{d-3}}+\frac{2}{(d-3)(d-2)}\frac{\alpha^2Q^2}{r^{2(d-3)}}
 -\frac{4}{(d-5)}\frac{\alpha^2 w_0^4}{r^2},
\end{eqnarray}
and
\begin{eqnarray}
\label{V}
~V(r)=V_0-\frac{Q}{(d-3)r^{d-3}},
\end{eqnarray}
with $V_0$ a constant which is fixed by requiring that the electric potential vanish at the horizon.
Apart from $w_0$, this solutions possesees two more parameters: $M_0$ and $Q$, which fixes the mass
and the electric charge densities, respectively. 

This black brane possesses an horizon at $r=r_H>0$, where $N(r_H)=0$ (and $N'(r_H)\geq 0$).
The Hawking temperature $T_H$, the event horizon area density $A_H$,
the chemical potential $\Phi$  and the electric charge density $Q_e$  
 of this solution are
\begin{eqnarray}
\label{AHTH}
&&
T_H=\frac{1}{4\pi}
\left(
(d-1)\frac{r_H}{L^2}-\frac{2\alpha^2}{r_H}
\bigg( 
\frac{2w_0^4}{r^2_H}
+
\frac{1}{(d-2)}\frac{Q^2}{r_H^{2(d-3)}} 
\bigg)
\right),~~
A_H=r_H^{d-2},
\\
\nonumber
&&
\Phi=\frac{1}{d-3}\frac{Q}{r_H^{d-3}},~~Q_e=\frac{\alpha^2}{4\pi}Q.
\end{eqnarray} 

One can easily verify that the total mass of the solutions, as defined 
according to the counterterm prescription in  
\cite{Balasubramanian:1999re}, diverges for any (even) 
$d>4$ due to the slow decay
of the magnetic fields, despite the fact that the spacetime is still AAdS.
A finite mass density results when a boundary term
\begin{eqnarray}
\label{Ict-mat-gen}
I_{ct}^{(YM)}=
- 
\frac{1}{d-5}
 \int_{\partial {\cal M}} d^{d-1}x\sqrt{-h }\frac{L}{4 }
 ~ F_{ab}^{IJ}F^{IJ~ab} ~,
\end{eqnarray}
is included in (\ref{action}),
with $h_{ab}$ the boundary metric and $F_{ab}^{IJ}$
the gauge field on the boundary.
Then the boundary stress tensor
$T_{ab}=\frac{2}{\sqrt{-h}}\frac{\delta I}{\delta h^{ab}}$
acquires a supplementary contribution from (\ref{Ict-mat-gen}),
which leads to finite mass density\footnote{As usual in this context,
the mass is the charge associated
with the time-translation symmetry of the boundary metric.} 
\begin{eqnarray}
\label{mass}
M=\frac{(d-2)}{16 \pi G}M_0.
 \end{eqnarray}
Note that this relation holds also for the simplest case $d=4$,
in which case no matter counterterm is required. 

 One can 
see that the quantities (\ref{AHTH}), (\ref{mass})
verify the first law of thermodynamics (with a constant background magnetic field)
\begin{eqnarray}
\label{first-law}
dM=\frac{1}{4 G}T_H dA_H+\frac{1}{ G}\Phi dQ_e.
\end{eqnarray}
 
 %%%%%%%%%%%%%%%%%%%%%%%%%%%%%%%%%%%%%%%%%%%%%%%%%%%%%%%%%%%
 \setlength{\unitlength}{1cm}
\begin{picture}(8,6) 
\put(-1,0.0){\epsfig{file=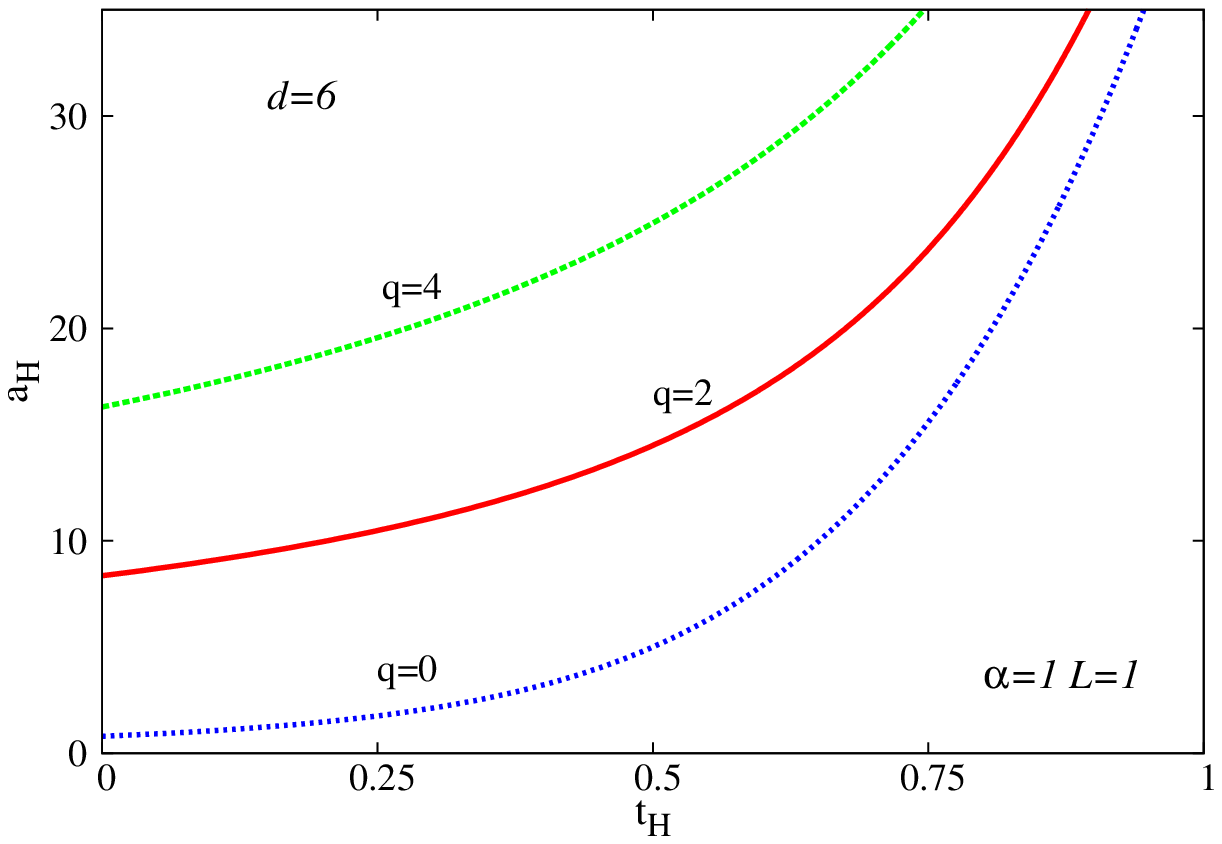,width=8.5cm}}
\put(8.2,0.0){\epsfig{file=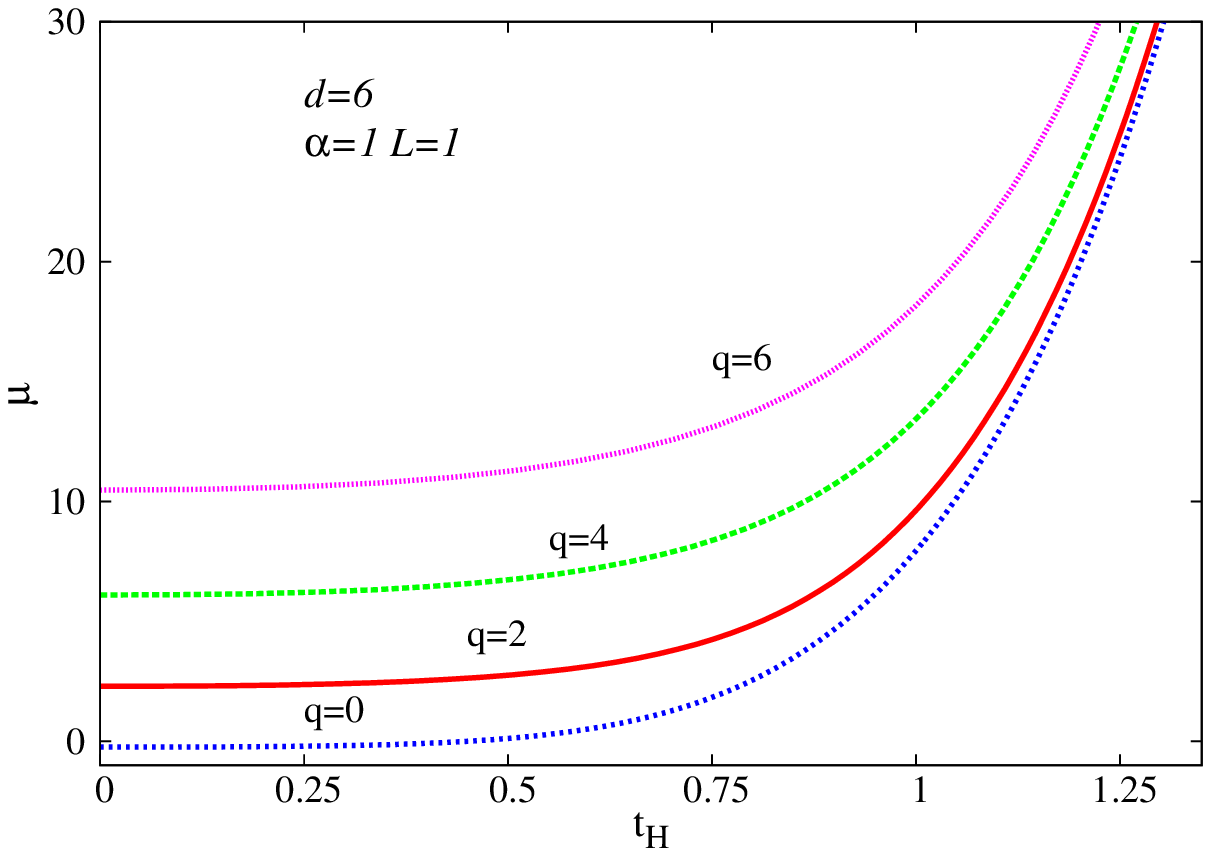,width=8.5cm}}
\end{picture}
\\
{\small {\bf Figure 1.} 
The reduced area $a_H$ and mass $\mu$ are shown as a function of reduced temperature $t_H$ for 
$d=6$ isotropic black branes in Einstein-Maxwell theory.
Here and in Figure 3 the quantities are scaled with respect to the magnetic field on the boundary.
}
\vspace{0.5cm}
%%%%%%%%%%%%%%%%%%%%%%%%%%%%%%%%%%%%%%%%%%%%%%%%%%%%%%%%%%%%  
 
In discussing the properties of these solutions (and their
non-Abelian generalizations),
it is convenient to work with quantities scaled 
with respect to the magnetic field in a two plane as fixed by the parameter
 $w_0$:
 \begin{eqnarray}
\label{scaled-quant}
a_H=\frac{A_H}{w_0^{ d-2 }},~~
t_H=\frac{T_H}{w_0},~~
\mu=\frac{G M}{w_0^{ d-1}},~~
q=\frac{Q_e}{w_0^{ d-2}}.~~
\end{eqnarray}
As seen in Figure 1,
the properties of the solutions with a 
background  magnetic field are not really sensitive to 
the presence of an electric charge
% (as encoded in the parameter $q$),
since the constant-$q$ curves preserve the $q=0$ shape,
which is approached asymptotically for large $t_H$.
These dyonic black branes  possess a regular extremal limit $T_H=0$,
with an $AdS_2\times R^{d-2}$ near horizon geometry.
An interesting feature is that the total mass of the $d>4$  solutions is
allowed to take negative values.
This can easily be seen in the extremal case, a limit which is approached for 
$Q=\frac{r_H^{d-2}}{\sqrt{2} \alpha L}\sqrt{(d-2)(d-1)}\sqrt{1-\frac{4\alpha^2 L^2 w_0^4}{(d-1) r_H^4}}$.
The extremal solutions have a mass
\begin{eqnarray}
\label{M-extremal}
M=\frac{(d-2)^2(d-1)}{8(d-3)\pi L^2}
\left(
1-\frac{4(d-4)\alpha^2 L^2 w_0^4}{(d-5)(d-2)(d-1) r_H^4} 
\right),
\end{eqnarray}
which becomes negative for
$
\frac{(d-5)(d-2) }{4(d-4)\alpha^2 L^2}<\frac{w_0^4}{r_H^4}\leq \frac{(d-1)}{4\alpha^2L^2}
$.
% some range of the ratio $w_0/r_H $.

%%%%%%%%%%%%%%%%%%%%%%%%%%%%%%%%%%%%%%%%%%%%%%%%%%%%%%%%%%%%%%%%%%%%%%%
 \noindent{\textbf{~~~Non-Abelian solutions.--~}}
 %%%%%%%%%%%%%%%%%%%%%%%%%%%%%%%%%%%%%%%%%%%%%%%%%%%%%%%%%%%%%%%%%%%%%%%
A simple nA ansatz leading to an isotropic line element can be constructed for
any $d\geq 4$, in terms of a magnetic potential, $w(r)$ and an electric one
%\footnote{
%The nonvanishing components of the field strength tensor are
%$
%F_{r i}^{IJ}=\frac{1}{\hat g}w'(r)\delta_{[i}^{~I}\delta_{d-1]}^{~J},
%$
%
%$
%F_{ij}^{IJ}=-\frac{1}{\hat g}w^2(r)\delta_{[i}^{~I}\delta_{j]}^{~J},
%$
%and 
%$
%F_{rt}^{IJ}=\frac{1}{\hat g}V'(r)\delta_{[d}^{~I}\delta_{d-1]}^{~J}. 
%$
%},  
$V(r)$ 
\begin{eqnarray}
\label{YM-ansatz}
A_i^{IJ}=\frac{w(r)}{\hat g}\delta_{[i}^{~I}\delta_{d-1]}^{~J},~~
A_r^{IJ}=0,~~A_t^{IJ}=\frac{V(r)}{\hat g}\delta_{[d}^{~I}\delta_{d+1]}^{~J}.
\end{eqnarray}
Unfortunately, no AAdS exact solutions
of the EYM equations seems to exist in this case.
However, the system possesses a simple globally regular Lifshitz-type configuration with 
\begin{eqnarray}
\label{exact-sol1}
ds^2=c_1\frac{dr^2}{r^2}+c_2r^2 d\Sigma_{d-2}^2-r^{2z}dt^2,
~~
 w(r)=u_0 r,
 ~~~
 V(r)=0,
  \end{eqnarray}
 where
\begin{eqnarray}
\label{exact-sol2}
c_1=\frac{4\alpha^2}{(d-2)p^2},~~c_2=\frac{2\alpha^2 
\left(
2(d-3)-(d-2)p^2)
\right)}{(d-2)^2p^2}u_0^2,~~
z=\frac{(d-3)((d-2)p^2+2)}{2(d-3)-(d-2)p^2}>1,
  \end{eqnarray}
here $u_0 \neq 0$ is an arbitrary constant while 
 $p$ is a parameter related  to the cosmological constant by 
 \begin{eqnarray}
\label{exact-sol3}
\nonumber
\Lambda=-\frac{(d-2)p^2}{2\alpha^2((d-2)q^2-2(d-3))}
\left(
(d-2)p^4+(d-2)(d-3)(d(d-6)+4)p^2+4(d-3)^2(d-1)
\right),
 \end{eqnarray}
and obeying the condition $p<\sqrt{2(d-3)/(d-2)}$. 
The solution (\ref{exact-sol1}) possesses the Lifshitz scaling symmetry
% \begin{eqnarray}
%\label{scaling-L}
$
t\to \lambda^z t,~~ x^i \to \lambda x^i,~~r\to r/\lambda
$
 %  \end{eqnarray},
 and generalizes the $d=4$ EYM solution of Ref. \cite{Devecioglu:2014iia} to higher dimensions.
 As discussed there, in this case the field equations possess
 black brane solutions with a regular horizon
 approaching the background (\ref{exact-sol1}) as $r\to\infty$.
We expect the existence of similar black brane solutions 
for $d>4$ as well.

Returning to the case of solutions  with AdS asymptotics,
it turns out   convenient for the  numerical construction
to choose a metric ansatz of the form
 \begin{eqnarray}
 \label{metric-new}
g_{rr}=\frac{1}{N(r)},~~g_{\Sigma\Sigma}=r^2,~~g_{tt}=-N(r)\sigma^2(r),~~
{\rm with}~~~
 N(r)=\frac{r^2}{L^2}-\frac{m(r)}{r^{d-3}}.
\end{eqnarray}

%%%%%%%%%%%%%%%%%%%%%%%%%%%%%%%%%%%%%%%%%%%%%%%%%%%%%%%%%%%%%%%%%%%%%%%%%%%%%%%
%\subsection{The ansatz and equations}
%%%%%%%%%%%%%%%%%%%%%%%%%%%%%%%%%%%%%%%%%%%%%%%%%%%%%%%%%%%%%%%%%%%%%%%%%%%%%%%

Inserting this ansatz into the Einstein and Yang-Mills equations yields four
 equations of motion\footnote{One extra equation containing the second derivatives of 
 the metric functions $m(r)$, $\sigma(r)$ is also found.
 However, one can show that this constraint equation is implicitly satisfied for the
 set of boundary conditions chosen.} for  $m(r),~\sigma(r),~w(r)$ and $V(r)$
 (a prime denotes $\frac{d}{d r}$):
\begin{eqnarray}
\nonumber
&&
m'=2\alpha^2 r^{d-4} 
\left (
\frac{1}{d-2}\frac{r^2V'^2}{\sigma^2}+Nw'^2+\frac{(d-3) }{2r^2}w^4
\right),
\\
\label{eqs}
&&
\sigma'=\frac{2\alpha^2}{r}  \sigma w'^2 ,
\\
\nonumber
&&
w''+ 
\left (
\frac{d-4}{r}+\frac{N'}{N}+\frac{\sigma'}{\sigma}
\right)w'
-(d-3)\frac{w^3}{r^2N}
=0,
\\
\nonumber
&&
%\left(\frac{r^{d-2}}{\sigma}V' \right)'=0,
V''+
\left(
\frac{d-2}{r}-\frac{\sigma'}{\sigma}
\right)
V'=0
\end{eqnarray}
 The last equation above implies the existence of the first integral
\begin{eqnarray}
\label{first-int}
V'=\sigma \frac{Q}{r^{d-2}},
 \end{eqnarray} 
 with $Q$ a constant fixing the electric charge of the solutions.
 
The equations of motion (\ref{eqs})
are invariant under three scaling transformations (invariant quantities are not shown):
\begin{eqnarray}
(I) 
&& \sigma\rightarrow \lambda\sigma, \qquad V\rightarrow \lambda V,  
\nonumber 
\\ 
(II) && r\rightarrow \lambda r\,, \quad m\rightarrow \lambda^{d-1} m \,, \quad w\rightarrow \lambda w \,, \quad V\rightarrow \lambda V, \,   
\label{scale}
\\  
(III) && r\rightarrow \lambda r\,,\quad m\rightarrow \lambda^{d-3} m\,, \quad L\rightarrow \lambda L\,,\quad V \rightarrow \frac{V}{\lambda}\,, \quad \alpha \rightarrow \lambda \alpha,  
\nonumber
\end{eqnarray} 
where  $\lambda$ represents the positive (real) scaling parameter.
Using $(I)$,   we  set the boundary values of the metric function $\sigma(r)$   to one, 
so that the metric will be asymptotically (locally) AdS. 
We are free to use $(II)$ to set 
the asymptotic value of the magnetic potential $w(r)$
to an arbitrary (non-vanishing) value
(equivalently, one can use this symmetry to 
fix the value of the electric charge or the horizon radius of the solution, say $r_H$).
%(as fixed by $Q$),
Finally, the symmetry  $(III)$ 
can be used to fix the value of the AdS radius $L$ or the value of the coupling constant $\alpha$;
for most of the work in this paper we
 set $\alpha=1$ (thus we treat $L$ as an input parameter).
% the AdS radius $L$ to one.

% $Q_e=\frac{\alpha^2}{4\pi}Q$.
 
Denoting the position of the horizon of the  black brane solutions by $r_H$, we have to impose 
$N(r_H)=0$ (and  $N'(r_H)\geq 0$) while the other metric functions  stay strictly positive.
A nonextremal solution
% (the only case discussed here) 
has the following expression near the event horizon:
\begin{eqnarray} 
 \nonumber
m(r)&=&\frac{r_H^{d-1}}{2L^2}+m'(r_H)(r-r_H)+O(r-r_H)^2, ~~
 \sigma(r)=\sigma_H+\sigma'(r_H)(r-r_H)+O(r-r_H)^2, 
 \\
\label{expansion}
w(r)&=&w_H+w'(r_H)(r-r_H)+O(r-r_H)^2, 
~~
V(r)=V'(r_H)(r-r_H)+O(r-r_H)^2,
\end{eqnarray}
where
\begin{eqnarray} 
\nonumber
&&m'(r_H)=\frac{2\alpha^2Q^2 }{(d-2)r_H^{d-2}}+\alpha^2(d-3)\frac{w_H^4}{r_H^{d-6}},
~~
w'(r_H)=\frac{(d-3)L^2 w_H^3}
{r_H^3\bigg(d-1-\frac{\alpha^2 L^2}{r_H^4} (\frac{2Q^2 }{(d-2)r_H^{2(d-4)}}+(d-3)w_H^4)\bigg ) } ,
\\
 &&V'(r_H)=\frac{Q}{r_H^{d-3}},
~~
 \sigma'(r_h)=-\frac{2\alpha^2(d-3)^2 L^4 \sigma_H w_H^6}{r_H^7
 \bigg(d-1-\frac{\alpha^2 L^2}{r_H^4} (\frac{2Q^2 }{(d-2)r_H^{2(d-4)}}+(d-3)w_H^4)\bigg )},
\end{eqnarray} 
with $w_H $ and $\sigma_H$ arbitrary constants.

The AdS boundary is reached as $r\to \infty$.
We are interested in configurations with $w(r)\to w_0 \neq 0$,  
  such that the nagnetic field on the boundary is  nonvanishing,
$
F_{ij}^{IJ}=-\frac{1}{\hat g}w_0^2\delta_{[i}^{~I}\delta_{j]}^{~J}.
$
A straightforward but cumbersome computation leads
  to the following general asymptotic expression of the solutions as $r\to \infty$
(note the presence of $\log$ terms for an odd value 
of the spacetime dimension):
\begin{eqnarray}
\nonumber
\label{asm}
&&m(r)=M_0+\frac{\alpha^2 L^{d-5}w_0^{d-1}}{d-2}\log(\frac{r}{L})
\left(
6\delta_{d,5}
-40\delta_{d,7}
+\frac{567}{4}\delta_{d,9}+\dots
\right)
\\
\nonumber
&&{~~~~~~~~~~}+\frac{\alpha^2(d-3)}{(d-5)}w_0^4r^{d-5}\mathsf{\Theta }\left( d-6\right)
-\frac{2\alpha^2L^2(d-3)^2(d-6)}{(d-5)^2(d-7) }w_0^6 r^{d-7}\mathsf{\Theta }\left( d-8\right)+\dots,
\\
\label{asympt}
&&\sigma(r)=1-\frac{4}{3}\alpha^2  w_0^6 \log^2(\frac{r}{L})\frac{L^4}{r^6}\delta_{d,5} -\frac{\alpha^2(d-3)^2L^4w_0^6}{3(d-5)^2r^6}\mathsf{\Theta }\left( d-6\right)
+\dots~~,
\\
\nonumber
\label{asw}
&&w(r)=w_0+\frac{J}{r^{d-3}}+\frac{w_0^{d-2}L^{d-3}}{r^{d-3}}\log(\frac{r}{L})
\left(
- \delta_{d,5}
+3\delta_{d,7}
-\frac{27}{4}\delta_{d,9}+
\dots\right)
\\\nonumber
&&{~~~~~~~~~~}-\frac{d-3}{d-5}\frac{w_0^3L^2}{2r^2} \mathsf{\Theta }\left( d-6\right)
+\frac{3(d-3)^2}{8(d-5)(d-7)}\frac{w_0^5L^4}{ r^4}\mathsf{\Theta }\left( d-8\right)+\dots,
\\
\label{asV}
\nonumber
&&V(r)=V_0-\frac{Q}{r^{d-3}}+\dots,
 \end{eqnarray}

The series 
truncates for any fixed dimension, with new terms entering at every 
new even value of $d$, as denoted by the step-function ($\mathsf{\Theta 
}\left( x\right) =1$ provided $x\geq 0$, and vanishes otherwise).
 The constants $w _0$, $M_0$, $V_0$ and $J$ in the above expressions are free parameters which are fixed by numerics.
% In what follows, we shall present arguments that $M_0$ fixes the mass of the $d=5,6$ configurations.
% At the same time, no global charge is associated with the constants $w_0,~C_1$
% in the asymptotics of the magnetic gauge potential.

As in the Abelian case,  we expect the parameter $M_0$ to
encode  the mass density of the solutions, which is still given by (\ref{mass}).
However, a rigorous proof of this statement is rather difficult,
due to the complicated asymptotic behaviour of the metric functions.
For $d=5$, a regularized boundary energy-momentum tensor and mass are found by 
including in (\ref{action})
 the following matter counterterm
\begin{eqnarray}
\label{Ict-mat5}
I_{ct}^{YM}=
- 
\log(\frac{r}{L})
 \int_{\partial {\cal M} }d^{4}x\sqrt{-h }\frac{L}{4 }
 ~ F_{ab}^{IJ}F^{IJ~ab} .
\end{eqnarray}
We have found that the 
boundary counterterm (\ref{Ict-mat-gen})
regularizes also the mass of the $d=6$ solutions.
%leading to the expression (\ref{mass}).
In both cases, this results in the expression (\ref{mass})
for the mass density of the black branes.
(Note that (\ref{mass}) holds also for $d=4$,
in which case no matter counterterm is necessary.)
However, the above simple counterterm fails to regularize
 all divergencies in the expression of $M$ for $d>6$.
Thus a more general matter counterterm than (\ref{Ict-mat-gen}) is required in the $d>6$ case. 
We find that for any $d\geq 4$, the mass of the solutions computed
by integrating the first law equation (\ref{first-law}),
coincides with the relation (\ref{mass}) with good accuracy.

Other quantities which enter the thermodynamics of the solutions are given by 
\begin{eqnarray}
\label{quant}
A_H=r_H^{d-3},~~T_H=\frac{1}{4\pi}N'(r_H)\sigma(r_H),~~\Phi=V_0,~~Q_e=\frac{\alpha^2}{4\pi}Q.
\end{eqnarray}

%Numerical s
Solutions interpolating between the near horizon expansion (\ref{expansion})
and the far field asymptotics (\ref{asympt})
are constructed numerically,
% by  
using a standard Runge-Kutta
ordinary  differential equations solver. In this approach we 
evaluate the initial conditions at   $r=r_H+10^{-5}$, for global tolerance $10^{-14}$,
adjusting  for shooting parameters and integrating  towards  $r\to\infty$
(thus we have restricted our study to the region outside the event horizon).
The equations were integrated for all values of $d$ between four and ten; thus similar solutions are expected to exist
for any value of $d$.

For a given $d$, we have considered a range of values for $(r_H,~w _{H},~Q)$, the parameters
$\sigma_H$ and $M_0,~V_0,~J$ resulting from the numerical output.
Since the equations (\ref{eqs})  are invariant under the
transformation $w \rightarrow - w $, only values of
$w _{H}>0$ are considered. 
Also, we have studied mainly the case where the AdS length scale is set to one, $L=1$.
The profile of a typical $d=6$ non-Abelian  solution   
 is shown in Figure 2. 
 (There we have displayed also the  mass function density $m(r)_{reg}$
regularized via the counterterm (\ref{Ict-mat-gen}).)

%%%%%%%%%%%%%%%%%%%%%%%%%%%%%%%%%%%%%%%%%%%%%%%%%%%%%%%%%%%
 \setlength{\unitlength}{1cm}
\begin{picture}(8,6) 
\put( 3,0.0){\epsfig{file=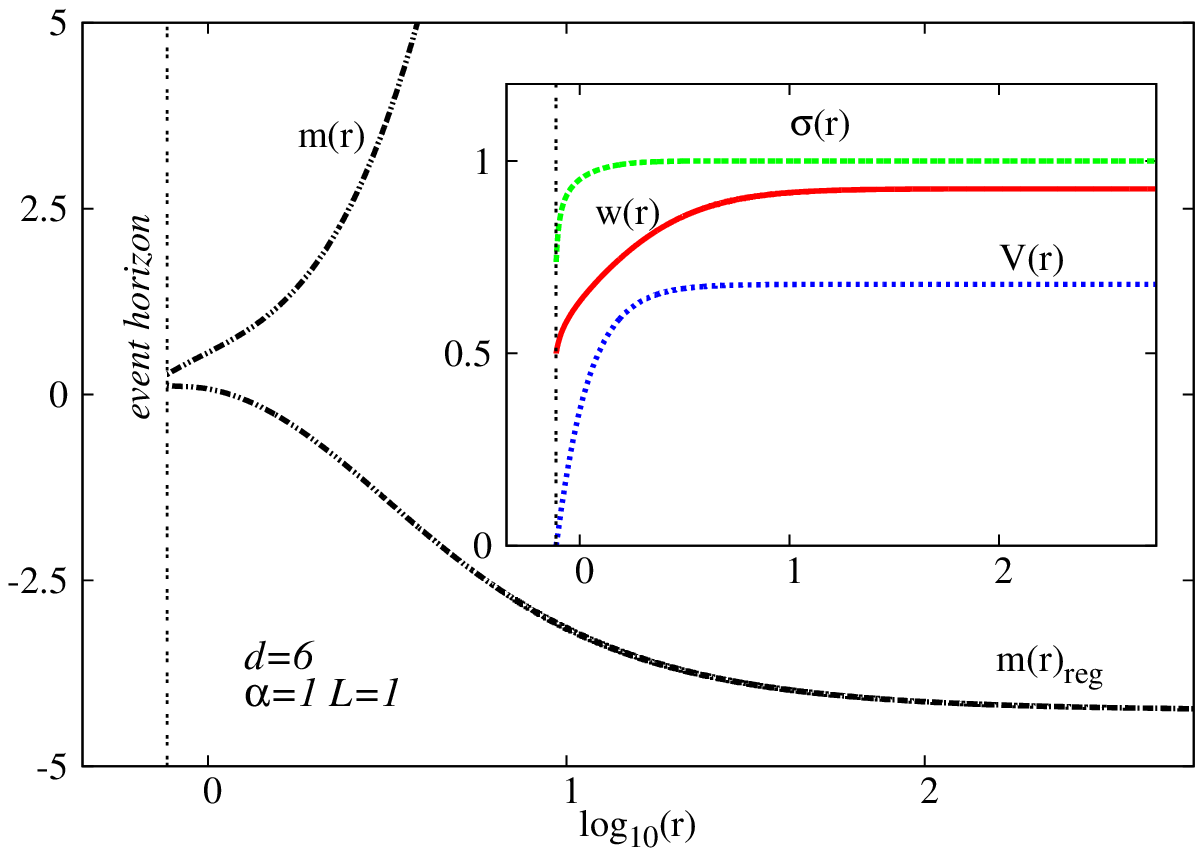,width=9.cm}}
\end{picture}
\\
{\small {\bf Figure 2.} 
 The  profiles  of a typical $d=6$ Einstein--Yang-Mills isotropic black brane solution
 are shown as a functions of the radial coordinate $r$.
 }
\vspace{0.5cm}
%%%%%%%%%%%%%%%%%%%%%%%%%%%%%%%%%%%%%%%%%%%%%%%%%%%%%%%%%%%%

 %%%%%%%%%%%%%%%%%%%%%%%%%%%%%%%%%%%%%%%%%%%%%%%%%%%%%%%%%%%
 \setlength{\unitlength}{1cm}
\begin{picture}(8,6) 
\put(-1,0.0){\epsfig{file=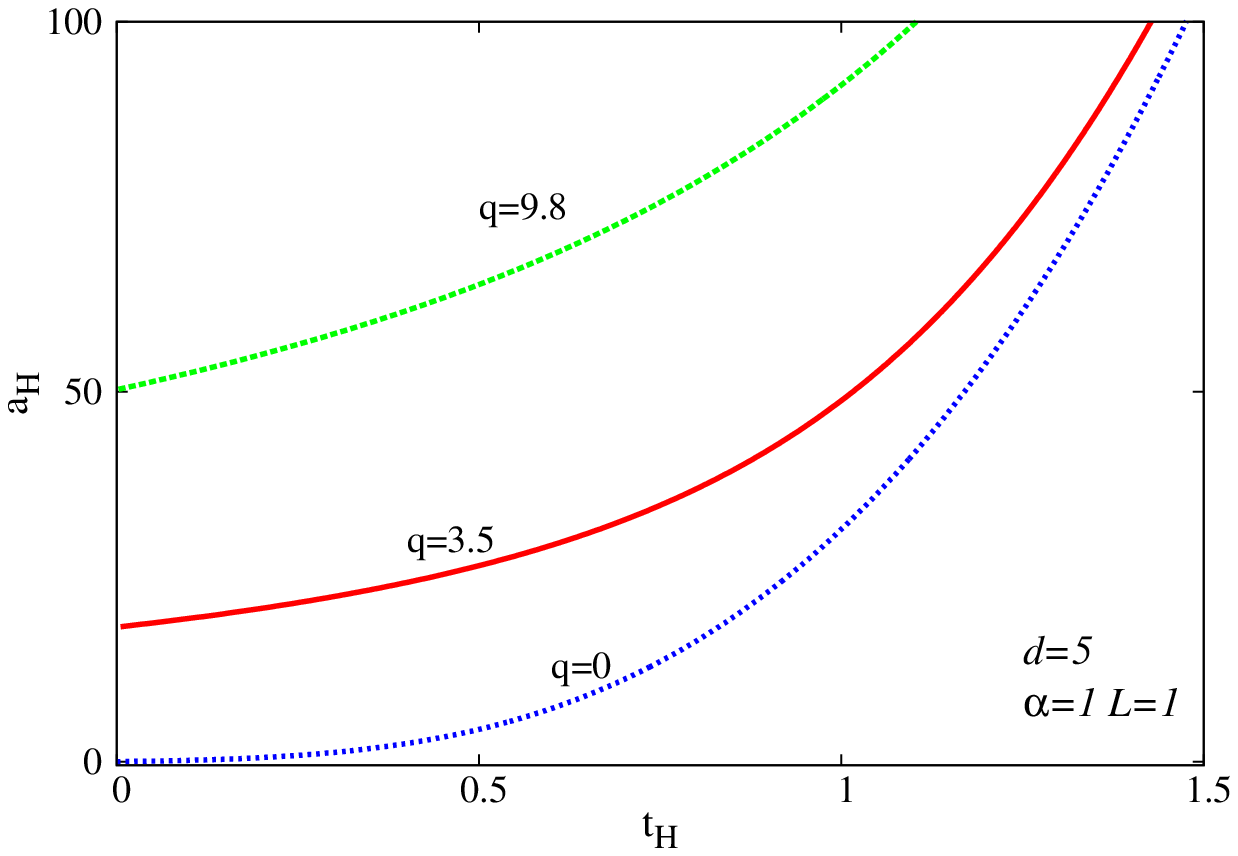,width=8.5cm}}
\put(8.2,0.0){\epsfig{file=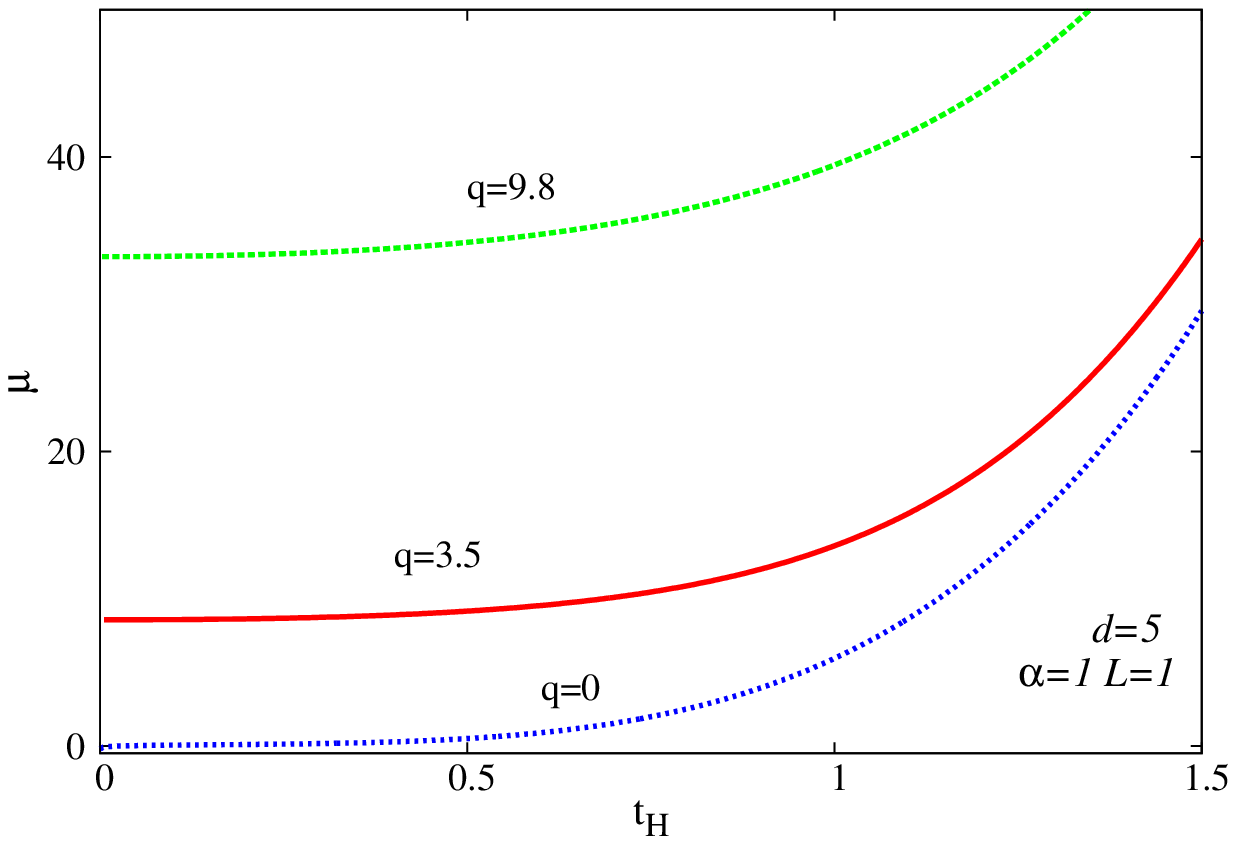,width=8.5cm}}
\end{picture}
\\
{\small {\bf Figure 3.} 
The reduced area $a_H$ and mass $\mu$ are shown as functions of the reduced temperature $t_H$ for 
$d=5$ isotropic black branes in Einstein--Yang-Mills theory.}
\vspace{0.5cm}
%%%%%%%%%%%%%%%%%%%%%%%%%%%%%%%%%%%%%%%%%%%%%%%%%%%%%%%%%%%%

We have found that 
the nA solutions share most of the basic properties of the Einstein-Maxwell
configurations discussed above.
In particular, the presence of an electric charge does not 
change qualitatively the general picture. Also, a
 number of basic features of these black holes are similar to those of the
known $d=4$ (purely magnetic) configurations in \cite{VanderBij:2001ia}.
This can be understood by noticing that, for our choice of the ansatz, 
the magnetic and electric 
potentials interact only via the spacetime geometry.
As a result,
these black branes can be
thought of as  nonlinear superpositions  of purely electric 
Reissner-Nordstr\"om-AdS solutions ($i.e.$ the limit $w_0=0$ in 
(\ref{Maxwell-ansatz}),
(\ref{metric1}))  
and purely magnetic nA configurations\footnote{One interesting feature  is the absence of  solutions with nodes of the magnetic
potential. 
This can be analytically proven by 
integrating the equation for $w$, 
$(N \sigma r^{d-4}w' )'= (d-3)w^3 \sigma r^{d-6} $, between $r_H$ and some $r$; 
%thus we 
obtaining $w'>0$ for every $r>r_H$.
In a similar way, one can prove that the metric function $\sigma(r)$
monotonically increases towards its asymptotic value.
} with $V(r)=0$.
This can be seen in Figure 3, where 
we plot the event horizon area and the mass
 of $d=5$  solutions, for several (fixed) values of the electric charge;
note that in that plot the quantities are normalized $w.r.t.$ to the magnetic field on the boundary, 
as defined by (\ref{scaled-quant}), which remain invariant under the transformation (ii) in (\ref{scale}).
One can easily see that the corresponding $q=0$ curves are generic.
Also, as in the Abelian case, we have noticed the existence of $d>5$
solutions with a negative total mass, $M_0<0$, see Figure 2 (solutions with $M_0=0$ do also exist).

However, the limiting behaviour of the EYM solution is
very different from the Abelian case, the limit $T_H\to 0$
being singular this time.
This can be understood by noticing that the non-linearity of the 
YM equation for the magnetic potential implies the
absence of a $AdS_2\times R^{d-2}$
near horizon geometry as a solution of the field equations.

%%%%%%%%%%%%%%%%%%%%%%%%%%%%%%%%%%%%%%%%%%%%%%%%%%%%%%%%%%%%%%%%%%%%%%%
 \noindent{\textbf{~~~Non-Abelian black branes
   in odd dimensions with a Chern-Simons term.--~}}
 %%%%%%%%%%%%%%%%%%%%%%%%%%%%%%%%%%%%%%%%%%%%%%%%%%%%%%%%%%%%%%%%%%%%%%%
 In odd spacetime dimensions, the usual gauge field
action can be augmented
%  by adding
with a Chern-Simons
(CS) term.
% in the action (\ref{action}).
%A very different picture is found when turning on a Chern-Simons  (CS) term 
Such a term  typically enters the action of gauged supergravities, the case of 
${\cal N}=8,~d=5$ model with a gauge group $SO(6)$, being perhaps the most interesting\footnote{
Note that a simple EYMCS theory does not seem to 
correspond to  a consistent truncation of any supergravity model.
However, we expect that the basic properties of our
solutions would hold also in that case
(see the Ref. \cite{Brihaye:2012nt}
for a study of nA in 
 of the ${\cal{N}}=4^+$, $d=5$ gauged supergravity model, which contains also a CS term.)
}.

 The  expression of the CS Lagrangean for the case 
 $d=5$  discussed in what follows, is\footnote{The explicit expression of the CS Lagrangean for 
 $d=7,~9$
 can be found $e.g.$ in Ref. \cite{Brihaye:2011nr}.}
\begin{eqnarray}
L_{CS}=\kappa  \ep_{I_1\cdots I_6} 
\Big(F^{I_1 I_2} \wedge  F^{I_3 I_4}\wedge  A^{I_5 I_6} 
  -
\hat g F^{I_1 I_2}\wedge  A^{I_3 I_4}\wedge  A^{I_5 J} \wedge A^{J I_6}
\\
\nonumber
 +\frac{2}{5} \hat g^2  A^{I_1 I_2}\wedge  A^{I_3 J}\wedge  A^{J I_4}\wedge 
A^{I_5 K} \wedge A^{K I_6} 
\Big),
\end{eqnarray}
with $\kappa$ an arbitrary parameter, the CS coupling constant\footnote{The value of $\kappa$
is fixed in supersymmetric theories, but in this work
we treat $\kappa$ as a free input parameter.}.

One can easily show that the Abelian configuration (\ref{metric1}) still remains a solution
 in the presence of a CS term\footnote{Note the situation changes 
 for $anisotropic$ dyonic Abelian black branes, 
 in which case the inclusion of a U(1) CS term leads to variety of new interesting properties,
 see $e.g.$ \cite{D'Hoker:2009bc}.
 };
however, the situation is different for non-Abelian fields. 
 These solutions can be studied within the same ansatz (\ref{YM-ansatz}), (\ref{metric-new});
the equations for metric functions $m(r)$, $\sigma(r)$ are still valid,
since the CS term does not contribute to the energy-momentum tensor,
while the equations for the gauge potentials
contain new terms encoding a direct interaction between magnetic
and electric potentials:
\begin{eqnarray}
\label{eqs-CS1}
&&
w''+ 
\left (
\frac{d-4}{r}+\frac{N'}{N}+\frac{\sigma'}{\sigma}
\right)
w'
-(d-3)\frac{w^3}{r^2N}-\kappa\frac{(d^2-1)}{(d-2)}\frac{w^{d-3}V'}{N\sigma r^{d-4}}
=0,
\\
\label{eqs-CS2}
&&
%%%% \frac{r^{d-2}}{\sigma}
 V'=\frac{\sigma}{r^{d-2}}(Q+\kappa\frac{(d^2-1)}{(d-2)}w^{d-2}),
\end{eqnarray}
with $Q$ an integration constant.

By using a similar approach to that described above,
we have studied families of $d=5$ solutions of the EYM-CS model in a systematic way~\footnote{
%Several solutions have been constructed also for a number $d=7$ of spacetime dimensions, they sharing the basic properties
%of the $d=5$ configurations.
We expect the properties of the five dimensional solutions to be generic.
Indeed, this is suported by the preliminary results we have found for EYMCS solutions
in  $d=7$ spacetime dimensions.}.

%%%%%%%%%%%%%%%%%%%%%%%%%%%%%%%%%%%%%%%%%%%%%%%%%%%%%%%%%%%
 \setlength{\unitlength}{1cm}
\begin{picture}(8,6)  
\put(-1,0.0){\epsfig{file=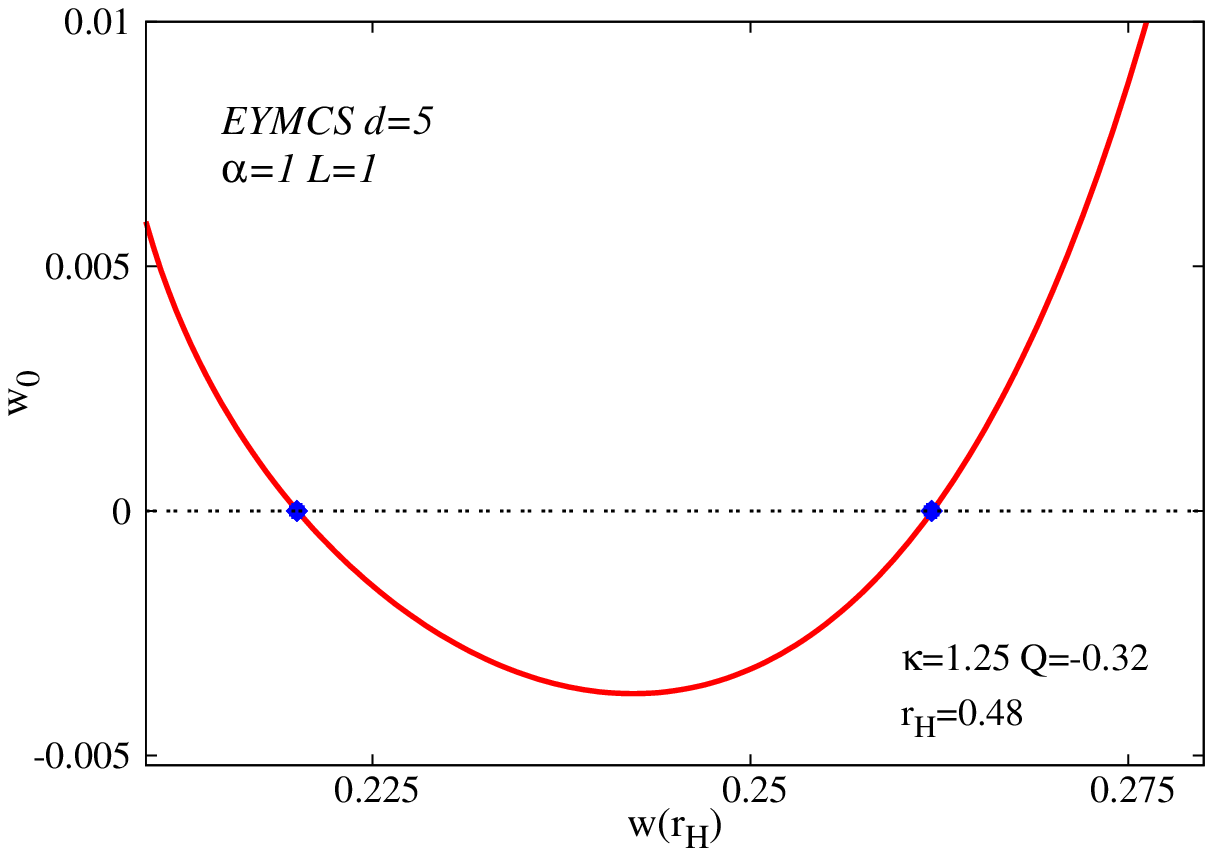,width=8.5cm}}
\put(8.2,0.0){\epsfig{file=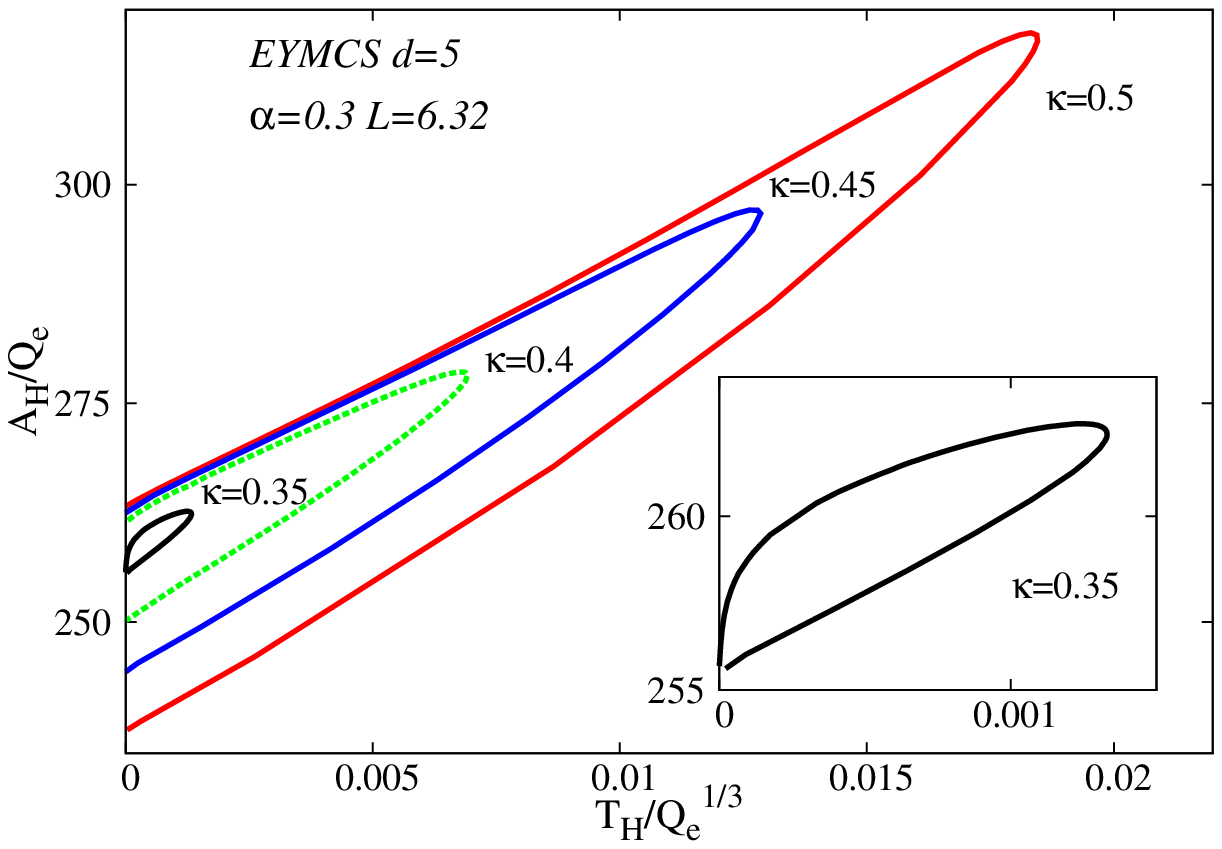,width=8.5cm}}

\end{picture}
\\
{\small {\bf Figure 4.} 
{\it Left:} The asymptotic value of the magnetic gauge potential $w_0$
is shown as a function of its value at the event horizon $w(r_H)$
for $d=5$ solutions in Einstein--Yang-Mills--Chern-Simons
theory.
{\it Right:}
The scaled horizon area $A_H$ is shown -for several values of $\kappa$- as a function of the scaled
temperature $T_H$ for $d=5$ Einstein--Yang-Mills--Chern-Simons solutions 
with a vanishing magnetic field on the boundary.
}
\vspace{0.5cm}
%%%%%%%%%%%%%%%%%%%%%%%%%%%%%%%%%%%%%%%%%%%%%%%%%%%%%%%%%%%%
 
The EYM-CS solutions possess a near horizon expansion similar to 
(\ref{expansion}), while their leading order terms in the far field expression is still given by (\ref{asympt}),
 $\kappa$ entering through the lower terms only.

We have found that all basic properties of the solutions without a CS term
are retained in this case. 
However, some new features occurs as well,
the most interesting being the 
 the existence of configurations
with $w(\infty)=0$.

For a special set of event horizon data, 
one finds solutions with vanishing magnetic field
on the AdS boundary (although $w(r)$
is nonzero in the bulk).
From (\ref{asympt}), this implies that in this case, as $r\to \infty$
the mass function $m(r)$ approaches a finite value. 
This feature is illustrated
in Figure 4 (left), where we plot $w_0$, the asymptotic value 
of the magnetic gauge potential $w$,
as a function of the value of the magnetic potential on the horizon 
 for fixed values of $\kappa,~Q,~r_H$ and $L$
(the special value of $w(r_H)$ which correspond to $w(\infty)=0)$
are marked with dots).

Naively, this resembles the solutions describing 
holographic $p-$wave superconductors and superfluids which have been extensively studied in recent years,
starting with the seminal work \cite{Gubser:2008zu}.
However, the overall picture is rather different
for the EYMCS solutions obtained here.
First, in contrast to the EYM solutions of Refs. 
\cite{Gubser:2008zu},
\cite{Manvelyan:2008sv},
\cite{Ammon:2009xh},
these  configurations
do not emerge as a perturbation of the RN-AdS Abelian solution\footnote{That is,
when
treating $w(r)$ as a small perturbation around 
the electrically charged RN-AdS black brane,
one finds that the 
solution of the YM-CS linearized equation (\ref{eqs-CS1}) 
possesses an essential logarithmic singularity at the horizon.}.
Second, the general pattern of  the EYMCS black branes
with a vanishing magnetic field on the boundary
is different from the one corresponding to nA configurations without a CS term.
For example, as seen in Figure 4 (right), the  EYMCS black branes with given $(\alpha,~\kappa)$
form two branches of solutions. These branches extend up to a maximal value of the Hawking temperature and horizon area, where they join
(note that the quantities plotted  are scale invariant under (ii) in (\ref{scale}) 
by an appropriate combination with the electric charge).

Interestingly (and in strong contrast to the pure EYM case discussed above),
the limit $T_H\to 0$
corresponds to extremal solutions with a regular horizon.
Such configurations possess an $AdS_2\times R^3$ near horizon geometry, with\footnote{This configuration can be generalized
for any (odd) $d\geq 5$; however, the relations are much more complicated in the general case.}
%EYMCS equations
\begin{eqnarray}
ds^2=v_1(\frac{dr^2}{r^2}-r^2 dt^2)+v_2 d\Sigma_3^2,
~~{\rm and}~~
w(r)=w_0,~~V(r)=q r,
\end{eqnarray}
(where the redefinition $r-r_H \to  r$ is implicitely assumed) and  
\begin{eqnarray}
v_1=\frac{2}{3}
\left(
\frac{8}{L^2}-\frac{\alpha^2 w_0^2}{16 \kappa^2(Q+8\kappa w_0^3)^2} 
\right)^{-1},~~
v_2=-\frac{4\kappa (Q+8\kappa w_0^3)}{w_0},
~~{\rm and}~~~q=\frac{v_1(Q+8\kappa w_0^3)}{v_2^{3/2}}.
\end{eqnarray}
 Given $\kappa$,~$\alpha$ and $L$, this configuration possesses
 one single free parameter, the constants $Q$, $w_0$
 satisfying the algebraic equation
 \begin{eqnarray}
512\kappa^2 (Q+8\kappa w_0^3)^2+\alpha^2 L^2 w_0^3(Q-4\kappa w_0^3)=0.
\end{eqnarray}

%%However,
We note that the overall picture possesses a nontrivial dependence 
on the value of the CS coupling constant,
with the existence of a minimal value of $\kappa$
allowing for a vanishing magnetic field
on the boundary.
We hope to return elsewhere with a systematic study of the EYMCS configurations, in a more 
general context.

%%%%%%%%%%%%%%%%%%%%%%%%%%%%%%%%%%%%%%%%%%%%%%%%%%%%%%%%%%%%%%%%%%%%%%%
 \noindent{\textbf{~~~Conclusions.--~}}
 %%%%%%%%%%%%%%%%%%%%%%%%%%%%%%%%%%%%%%%%%%%%%%%%%%%%%%%%%%%%%%%%%%%%%%%
 %
In this work we have constructed isotropic black branes in an $AdS_{d}$ background possessing
both electric and magnetic $SO(d+1)$ non-Abelian fields. The solutions were obtained by using a combination of
analytical and numerical methods.
Several basic properties of these solutions   in $d>4$  can hardly be anticipated from the
study of their four dimensional counterparts.
For example, the magnetic field of the EYM solutions does not vanish asymptotically.
%; thus 
As a result their mass -defined in the usual way-
always diverges.
% for $d>4$
However, solutions with a finite mass exist - in odd spacetime dimensions -
when supplementing the action by a Chern-Simons term.

There are various possible  natural extensions of this work.
Perhaps the most interesting one would be to study the transport properties of these solutions. 
%The i
Investigation of the thermodynamics of the black branes is another important problem.
Here we mention only that the heat capacity is always positive
for the EYM black holes in a canonical ensemble.
As a result, these configurations
are always thermodynamically locally stable, a feature shared with the vacuum solutions.
Finally, note that the YM ansatz used in this work is
not the most general one leading to an isotropic black brane;  for instance
the components of the connection (\ref{YM-ansatz}) take their values in the algebra 
of $SO(d - 1)\times U(1)$ and not in the full algebra of $SO(d+1)$.
The fully $SO(d+1)$ YM ansatz can be written in
terms of two magnetic potentials and two electric potentials,
and is expected to lead to a more complicated picture\footnote{The $d=5$ EYMCS counterpart of these configurations 
with a spherical horizon topology have been studied in 
\cite{Brihaye:2009cc},
\cite{Brihaye:2011dy};
see also 
\cite{Brihaye:2010wp},
\cite{Brihaye:2011nr} 
for the $\Lambda=0$ limit of these solutions.
}.

%Further analysis of this type of solutions and their role in low energy string theory remains an interesting issue to explore in the future.

 \vspace{1.cm}
{\bf\large Acknowledgements} \\
The work of  Y.B. was supported in part by an ARC contract AUWB-2010/15-UMONS-1. 
E.R. gratefully acknowledges support from the FCT-IF programme and
CIDMA strategic project UID/MAT/04106/2013.

%%%%%%%%%%%%%%%%%%%%%%%%%%%%%%%%%%%%%%%%%%%%%%%%%%%%%%%%%%%%%%%%%%%%%%%%%%%%%%

  \end{document}